\documentclass[prl,twocolumn,aps,amssymb,showpacs,superscriptaddress,nofootinbib]{revtex4}
\usepackage{graphicx}

\begin{document}
%
\preprint{IMSc/2005/04/10}

\title{Large volume quantum correction in
loop quantum cosmology: Graviton illusion?}

\author{Golam Mortuza Hossain}
\email{golam@imsc.res.in}
\affiliation{The Institute of Mathematical Sciences,
CIT Campus, Chennai-600 113, India.}

\begin{abstract}

The leading quantum correction to Einstein-Hilbert Hamiltonian
coming from large volume vacuum isotropic loop quantum cosmology,
is {\em independent} of quantization ambiguity parameters. It is
shown here that this correction can be viewed as finite volume
gravitational Casimir energy due to one-loop `graviton'
contributions. In vacuum case sub-leading quantum corrections and
in non-vacuum case even leading quantum correction depend on
ambiguity parameters. It may be recalled that these are in fact
analogous features of {\em perturbative} quantum gravity where it
is well-known that pure gravity (on-shell) is one-loop finite
whereas higher-loops contributions are not even renormalizable.
These features of the quantum corrections coming from
non-perturbative quantization, sheds a new light on a major open
issue; how to communicate between non-perturbative and
perturbative approaches of quantum gravity.

\end{abstract}


\maketitle


The Standard Model of particle physics, a description of matter
fields based on {\em perturbative} quantum field theory, has been
shown to be one of the most predictive physical theory ever
constructed. The physical predictions of this theory have been
verified experimentally with an outstanding accuracy.
Unfortunately, the techniques of perturbative quantum field
theory when applied to theory of gravity, fail quite miserably.
Thus, in a quest for a quantum theory of gravity one is being
compelled to take different courses. In the most popular
approach, known as {\em string theory} \cite{String}, one
postulates the basic constituents of the model to be one
dimensional objects. The other leading candidate theory of
quantum gravity is known as {\em loop quantum gravity} (LQG)
\cite{LQG} where one attempts to formulate a background
independent {\em non-perturbative} quantum theory of gravity.

Being very different in its formulation, a major issue that
haunts non-perturbative quantum gravity, is its relation to the
`low-energy' world \cite{Varadarajan}. In particular, how do the
results of non-perturbative approach compare with the results of
perturbative approach in the regime where perturbative approach
should be a good effective description. To explore this issue,
here we use the framework of {\em loop quantum cosmology} (LQC)
\cite{LQCall,Bohr}. It is a quantization of cosmological
(homogeneous) models using techniques of loop quantum gravity.
Thanks to its simplicity, it allows explicit calculation to study
its consequences. The loop quantum cosmology so far has lead to
several impressive results. It has been shown that loop quantum
cosmology cures the problem of classical singularity \cite{Sing},
along with quantum suppression of classical chaotic behaviour
near singularities in Bianchi-IX models \cite{BIX}. Further, it
has been shown that this model has a in-built generic phase of
inflation \cite{Inflation}. The corresponding power spectrum of
density perturbation contains a distinguishing feature
\cite{Hossain:PPS}. However, the issues related with physical
observables, external time evolution, physical Hilbert space are
still in nascent stage \cite{LQCall,OTH}. Nevertheless, it is
possible to derive an {\em effective} Hamiltonian using WKB
techniques \cite{EffectiveHam,Banerjee}.

\begin{figure}
\begin{center}
\includegraphics[width=6cm,height=4cm]{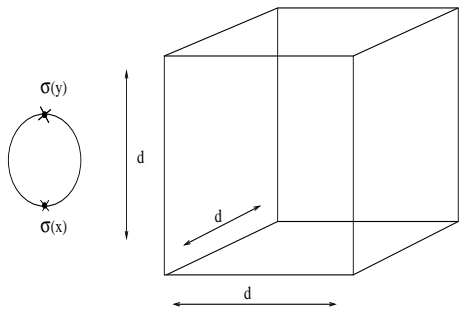}\\ 
\end{center}
\caption{The Feynman diagram with two point insertion is the
leading contributor to Casimir energy in small coupling limit.
A finite cell of universe with length of each side $d$.}
\label{fig}
\end{figure}

We consider here spatially flat isotropic loop quantum cosmology,
as we are interested in vacuum solution of it. The spatially
closed model does not have vacuum solution. In loop quantum
cosmology, a kinematical state is written as $|s\rangle = \sum
s_{\mu} |\mu\rangle$, where $|\mu\rangle$'s are eigenstates of
volume (densitized triad) operator. It is important to emphasize
the meaning of volume in this context. In particular, the volume
$V=\int d^3x {\sqrt{-g}}$ of the space is infinite, as it
is non-compact. To avoid this trivial divergence in loop quantum
cosmology, one considers the volume of a finite cell of universe
(see Fig \ref{fig}.) and studies its evolution.  This feature
plays the central role in the arguments presented here. In loop
quantum cosmology, the underlying (internal time) dynamics is
described by a {\em difference} equation. This discrete evolution
faithfully represents the underlying discrete geometry, a feature
of full theory of loop quantum gravity. In the effective
description of loop quantum cosmology, one tries to understand
the dynamics from a perspective based on continuum geometry. In
other words, one tries to approximate the fundamentally discrete
dynamics by a continuum dynamics. In this process, the discrete
dynamics effectively provides a new potential term in the
continuum description. This feature can be seen rather easily by
considering the gravitational term in the difference equation
$(A_{\mu + 4 \mu_0} s_{\mu + 4 \mu_0} - 2 A_{\mu} s_{\mu} +
A_{\mu - 4 \mu_0} s_{\mu - 4 \mu_0})$, where $A_{\mu} := {|\mu +
\mu_0|}^{3/2} - {|\mu-\mu_0|}^{3/2}$.  In deriving the effective
Hamiltonian, in first step one approximates the solution of the
difference equation $s_{\mu}$, by a smooth (differentiable)
function say $\psi(p:=\gamma \mu l_p^2/6)$. Using WKB
approximation, one derives a Hamilton-Jacobi equation from it.
The corresponding Hamiltonian then contains an effective
potential term, referred as {\em quantum geometry} potential
\cite{EffectiveHam} which is $(l_p^2~p_0^{-3}/288) (A_{p+4p_0} -
2A_{p} + A_{p-4p_0})$.  Thus, the quantum geometry potential term
arises when one tries to view a fundamentally discrete dynamics
through a continuum description. In effective description, the
quantum geometry potential leads to a generic bounce
\cite{GenBounce}. In large volume $V$($=p^{3/2}$, $p$ is
densitized triad) and small extrinsic curvature $K$ (conjugate
variable of $p$) regime, the gravitational part of the effective
Hamiltonian \cite{EffectiveHam,Banerjee} can be expanded as
\begin{eqnarray}
&\text{H}_{\text{grav}}^{\text{eff}}& =~
\text{H}_{\text{EH}} ~+~ l_p^2 \left[ -~ \frac{1}{24~ p^{3/2}}
~+~ \frac{2\mu_0^2\gamma^2}{9{\sqrt p}}~ \text{H}_{\text{EH}}^2
\right]\nonumber\\ 
&-& l_p^4 \left[
\frac{49~\mu_0^2\gamma^2}{864~p^2} \text{H}_{\text{EH}} \right] -
l_p^6 \left[ \frac{5~\mu_0^2\gamma^2}{768~p^{7/2}} + .. \right] +
.., 
\label{EffHam} 
\end{eqnarray}
where $\text{H}_{\text{EH}} = - (3/2\kappa) \sqrt{p}~ K^2$, is
the Einstein-Hilbert Hamiltonian for the homogeneous and
isotropic spacetime. In natural units ($c=\hbar=1$), $\kappa =
16\pi G = l_p^2$, is the gravitational coupling constant.
$\gamma$ is the Barbero-Immirzi parameter. $\mu_0$ here is viewed
as a quantization ambiguity parameter. $\mu_0$ appears as the
length of the edges, while expressing curvature tensor in terms
of holonomies around a square. It essentially plays the role of a
regulator\cite{Bohr}. Both of these parameters are generally
assumed to be order of unity numbers but there is no unique way
to fix their values within loop quantum cosmology itself. 
The accuracy of WKB approximation here
increases with increasing volume. So in large volume (scale
set by the step-size of the difference equation), the effective
Hamiltonian (\ref{EffHam}) is quite trustworthy.

Let's now consider the pure gravity case {\em i.e} without any
matter field. The Einstein-Hilbert Hamiltonian
$\text{H}_{\text{EH}}$ vanishes (on-shell) for pure gravity. It
is then clear from the expression (\ref{EffHam}) that leading
quantum correction is independent of the parameters $\mu_0$ and
$\gamma$. This quantum correction comes solely from the quantum
geometry potential. In the volume goes to infinity limit this
quantum correction vanishes. We will refer this term as
gravitational Casimir energy. Later we will show that this term
can indeed be viewed as gravitational Casimir energy due to the
finite volume of the system. We now re-write the term as 
\begin{equation}
\text{E}_{\text{Cas}}^{\text{grav}}~=~ -\frac{1}{24}
\frac{l_p^2}{d^3}~, 
\label{CasimirLQC}
\end{equation}
where the volume $d^3 = p^{3/2}$. 

Traditionally, one computes Casimir energy by computing the shift
of vacuum polarisation energy due to imposition of an external
boundary condition.  In particular, the {\em quantum
electrodynamic} Casimir energy between two conducting plates of
surface area $\tilde{A}$ and separated by a distance $d$, is
$-(\pi^2/240)(\tilde{A}/d^3)$.  Surprisingly, this expression
does not have explicit dependence on fine structure constant.
Although, {\em in principle} possible but {\em in reality} there
is no conductor that can enforce such boundary conditions for
modes of {\em all} wavelength. On the other hand, the
experimental result seems to agree with the traditional
expression extremely well \cite{Lamoreaux}. Thus, reconciling
these two facts may appear as a conceptually difficult task.
However in a recent approach of computing Casimir energy
\cite{Graham:Casimir,Jaffe:QV}, instead of imposing boundary
condition, one considers the plates as a classical static
background field. Then one introduces an interaction of the type
$-\mathcal{L}_{\text{int}} = \lambda~ \sigma(x) \phi^2(x)$ where
$\sigma(x)$ is classical (non-dynamical) background field and
$\phi(x)$ is the dynamical field whose vacuum fluctuations
contribute to Casimir energy. The background field $\sigma(x)$ is
represented by delta functions peaked around the positions of the
conducting plates. Using the techniques of perturbative quantum
field theory, one computes order by order contributions of this
interaction. It is possible to sum up all order contributions to
give a `close' form expression for the Casimir energy. In {\em
strong} coupling ($\lambda \rightarrow \infty $) limit, the
explicit dependence of coupling constant drops out from this
expression and it reduces exactly to the traditional expression
of Casimir energy. The real experimental system that one
considers for measuring Casimir force, the fine structure
constant effectively appears as strong coupling \cite{Jaffe:QV}.
Thus, the recent approach of computing Casimir energy addresses
the mentioned conceptual difficulty quite well. On the other hand
in {\em weak} coupling ($\lambda \rightarrow 0 $) limit, the
Casimir energy computed using this method, scales as $\sim
(\lambda^2/d)$. It is essentially the contribution from one-loop
diagram with two-point insertion.

The expression of effective Hamiltonian (\ref{EffHam}) is valid
in large volume ( $d^2 >> l_p^2$) regime. Naturally in the regime
of interest, the gravitational coupling constant $l_p$ is very
weak. We have already mentioned that in isotropic loop quantum
cosmology, one essentially studies the evolution of a finite cell
of the universe. So one can expect to get finite volume Casimir
energy due to quantum fluctuations of {\em geometry}. The
homogeneous and isotropic vacuum solution of Einstein equation is
Minkowski spacetime. Since we are considering vacuum isotropic
loop quantum cosmology, the system is essentially a finite patch
of the Minkowski spacetime. This feature makes it conceptually
easier to use the techniques of perturbative quantum field
theory.

To compute the gravitational Casimir energy for the system {\em
i.e.} a finite cell of the universe, here we use the recent
method. Essential difference in this case is that one should
consider the vacuum fluctuations of spin $2$ field. Also, instead
of one pair of boundary, here one has three pairs of boundary,
one pair in each spatial direction.  For simplicity, however we
will compute Casimir energy due to a massless spin $0$ field 
(massless Klein-Gordon field). The computational scheme can be
extended for the spin $2$ field as well. The result will differ
by a numerical factor of $2$ because of its two {\em helicities}.
So to study the {\em qualitative} behaviour (as it is not
expected to have {\em quantitative} match; in loop quantum
cosmology one consider only the {\em temporal} fluctuations of
geometry.  Imposition of high symmetry essentially freezes the
{\em spatial} fluctuations.), use of spin $0$ field suffices.

We consider the background field $\sigma(x)$ to be represented by
{\em three} dimensional delta functions, as there are boundaries
along all three spatial directions. We `normalize' non-dynamical
background field $\sigma(x)$ as
\begin{equation}
\int d^3\vec{x} \sigma(x) := \int d^3\vec{x} \frac{1}{2}[
{\bf \delta}^3( \vec{x} - \frac{\vec{d}}{2}) + 
{\bf \delta}^3(\vec{x} + \frac{\vec{d}}{2})] = 1 ~,
\label{BField}
\end{equation}
where $\vec{d}=\{d,d,d\}$. It is worth pointing out that a
different `normalization' essentially alters the coupling
constant $\lambda$. In this approach
\cite{Graham:Casimir,Jaffe:QV} Casimir energy is read off from
the one-loop {\em effective action} computed using the background
field method \cite{Peskin}. Naturally, the Casimir energy is
defined as
\begin{equation}
\int dt ~\text{E}_{\text{Cas}}
[\sigma ] := - \frac{i}{2}~ \text{log det}
\left[ - \frac{\delta^2 \mathcal{L}}{\delta\phi \delta\phi}
\right] ~,
\label{CasimirDef}
\end{equation}
where $\mathcal{L}$ is the full interacting Lagrangian. The
functional determinant can be expressed in terms of Feynman
diagrams. Being independent of $d$, the one-loop diagram with
one-point insertion does not contribute to Casimir {\em force}
which is a physically measurable quantity. The contribution from
one-loop diagram with two-point insertion (see Fig \ref{fig}.)
can be computed in a straightforward manner, leading to
\begin{equation}
\text{E}_{\text{Cas}}[\sigma] ~=~ - \frac{1}{24}
\frac{\alpha^2 \lambda^2}{d^3} ~,
\label{CasimirKG}
\end{equation}
where $\alpha^2 = {({\sqrt 3}(2\pi)^3)}^{-1}$. In the
calculation, the $d$ independent contributions (formally
divergent) have been dropped, as they do not contribute to
Casimir force.

Before we compare the expression (\ref{CasimirKG}) of Casimir
energy computed using perturbative quantum field theory, with the
expression (\ref{CasimirLQC}) extracted from isotropic loop
quantum cosmology, a caution is appropriate. In {\em strong}
coupling limit the method used here gives unambiguous expression
for Casimir energy as coupling constant dependence drops out.
However in {\em weak} coupling limit, it depends on the choice of
`normalization' (\ref{BField}) which needs to be provided from
outside. Also, the contribution from isotropic loop quantum
cosmology itself may not account for the full gravitational
Casimir energy. So here we restrict to the {\em qualitative}
comparison of these two expressions. Comparing the expressions
(\ref{CasimirLQC}) and (\ref{CasimirKG}), it is clear that the
expression (\ref{CasimirLQC}) can indeed be viewed as
contributions from virtual quantas whose coupling strength is
Planck constant $l_p$. So we refer these quantas as `gravitons'.
However, due to the `normalization' uncertainty involved in the
perturbative method used here, it is not yet possible to conclude
{\em definitively} about the spin degrees of freedom of these
quantas.

Let's now go back to the expression (\ref{EffHam}) of the
effective Hamiltonian. In vacuum case, the leading quantum
correction is unambiguous. However, the sub-leading corrections
depends on ambiguity parameters. With inclusion of matter, the
Einstein-Hilbert Hamiltonian $\text{H}_{\text{EH}}$ does not
vanish. Clearly, the leading quantum correction then also becomes
ambiguous. There will also be contributions from the direct
coupling of matters with gravity. It is important to observe that
these features of large volume quantum corrections coming from
non-perturbative quantization, in fact closely resemble the
features of {\em perturbative} quantum gravity. It is well-known
through the work of 't Hooft and Veltman \cite{tHooft:OL} that
pure gravity (on-shell) is one-loop (order $l_p^2$) finite. In
other words, one-loop contributions from perturbative quantum
gravity without matter, is unambiguous. However, higher-loops
contributions from pure gravity are not even renormalizable {\em
i.e.} it is not possible to obtain unambiguous results from such
computations. With inclusion of matter, perturbative quantum
gravity is not even one-loop renormalizable. Now, as we have
already mentioned that a severe criticism that often haunts the
advocates of non-perturbative quantum gravity, is its relation to
the `low energy' world. Although for symmetric models as shown
here, the quantum cosmology based on loop quantum gravity not
only reproduces the Einstein-Hilbert Hamiltonian as the leading
term but also its quantum corrections resemble the qualitative
features of perturbative quantum gravity in the regime where
the later should be a reasonable effective description.

In computing Casimir energy using perturbative quantum field
theory, the interaction term was introduced rather {\em by hand}.
We now argue that the form of the interaction used in the
calculation arises quite naturally. The gravitational Lagrangian
involves term of the form $g(x) g(x) \partial g(x) \partial
g(x)$. In the background field method of quantum field theory,
one expands the field $g(x)$ around a given classical background
say $\eta(x)$; $g(x) = \eta(x) + h(x)$, where $h(x)$ is the
fluctuating field. Inserting the decomposition into the
gravitational Lagrangian, it is easy to see that it contains a
term of the form $\sigma(x) h(x) h(x)$. The $\sigma(x) \sim
\partial \eta(x) \partial \eta(x)$, can indeed be treated as a
classical background field. For small extrinsic curvature regime,
one can simply consider background $\eta(x)$ as static while
computing perturbative corrections. For a finite cell of the
universe, the use of delta function potential peaked around the
boundaries is also well-motivated. For example, one crude way to
make the volume of flat space finite, is by multiplying the
metric component with a Heaviside step function say $\theta(
d^{\mu}/2 - |x^{\mu}|)$ where $d^\mu = \{\infty, \vec{d}\}$. It
is easy to see that the background field $\sigma(x)$ then
involves delta functions peaked around boundaries.  However, the
mentioned term need not be the only boundary interaction term
that can contribute to the Casimir energy. So, it is necessary to
perform a `first principle' computation of gravitational Casimir
energy for the system. It may also help to eventually settle the
issue of spin degrees of freedom through {\em quantitative}
comparison or at least to specify what to expect from a
computation using the full theory of loop quantum gravity.

To summarize, the leading quantum correction to Einstein-Hilbert
Hamiltonian coming from vacuum isotropic loop quantum cosmology
is {\em unambiguous} and can be viewed as gravitational Casimir
energy due to one-loop `graviton' contributions. However, based
on arguments presented here, it is not yet possible to conclude
{\em definitively} about the spin degrees of freedom of these
quantas. The sub-leading quantum corrections depend on
quantization ambiguity parameters. In non-vacuum case even
leading quantum correction depends on ambiguity parameters.
Importantly, these are analogous features of {\em perturbative}
quantum gravity. In other words, the quantum corrections coming
from loop quantum cosmology whose quantization relies on
non-perturbative techniques, closely resemble the qualitative
features of perturbative quantum gravity in the regime where the
later should be a reasonable effective description.

{\em Acknowledgements:}
I thank Ghanashyam Date, Romesh Kaul and Martin Bojowald for
careful reading and critical comments on the manuscript. It is a
pleasure to thank Ghanashyam Date, Romesh Kaul for helpful,
illuminating discussions. I thank Romesh Kaul for an
encouragement.


\end{document}